\documentclass[%
 reprint,
 superscriptaddress,
 amsmath,amssymb,
 aps,
 showkeys,
 titlepage,
prb,citeautoscript,
]{revtex4-1}

\usepackage[utf8]{inputenc}
\usepackage[english]{babel}
\usepackage{graphicx}
\usepackage{dcolumn}
\usepackage{bm}
\usepackage[mathlines]{lineno}
\usepackage[colorlinks = true,
            linkcolor = blue,
            urlcolor  = blue,
            citecolor = red,
            anchorcolor = blue]{hyperref}
\usepackage{color, soul}

\usepackage{comment}
\usepackage{natbib}
\makeatletter
\newcommand*{\rom}[1]{\expandafter\@slowromancap\romannumeral #1@}

\usepackage[version=4]{mhchem}
\newcommand{\code}[1]{\texttt{#1}}
\usepackage{chemformula}
\usepackage{threeparttable}
\usepackage{hhline}
\makeatother
\begin{document}


\title{Thermoelectric Transport in Weyl Semimetal \ce{BaMnSb2}: a First-Principles Study}

\author{Yubi Chen}
\affiliation{%
 Department of Physics, University of California, Santa Barbara, California 93106-9530, USA}%
 \affiliation{%
Department of Mechanical Engineering, University of California, Santa Barbara, CA 93106-5070, USA}%
\author{Rongying Jin}
\affiliation{%
 SmartState Center for Experimental Nanoscale Physics, Department of Physics and Astronomy, University of South Carolina, SC 29208, USA}%
\author{Bolin Liao}
\email{bliao@ucsb.edu}
\affiliation{%
Department of Mechanical Engineering, University of California, Santa Barbara, CA 93106-5070, USA}%
\author{Sai Mu}
\email{mus@mailbox.sc.edu}
\affiliation{%
 SmartState Center for Experimental Nanoscale Physics, Department of Physics and Astronomy, University of South Carolina, SC 29208, USA}%

\begin{abstract}
Topological materials are often associated with exceptional thermoelectric properties.
Orthorhombic \ce{BaMnSb2} is a topological semimetal consisting of alternating layers of Ba, Sb, and MnSb.
A recent experiment demonstrates that \ce{BaMnSb2} has a low thermal conductivity and modest thermopower, promising as a thermoelectric material. 
Through first-principles calculations with Coulomb repulsion and spin-orbit coupling included, we studied the electronic structure, phononic structure, and thermoelectric transport properties of \ce{BaMnSb2} in depth. 
We find that \ce{BaMnSb2} exhibits a low lattice thermal conductivity, owing to the scattering of the acoustic phonons with low-frequency optical modes.
Using the linearized Boltzmann transport theory with a constant relaxation time approximation, the thermopower is further calculated and an intriguing goniopolar transport behavior, which is associated with both $n$-type and $p$-type conduction along separate transport directions simultaneously, is observed. 
We propose that the figure of merit can be enhanced via doping in which electrical conductivity is decreased while the thermopower remains undiminished. 
\ce{BaMnSb2} is a potential platform for elucidating complex band structure effects and topological phenomena, paving the way to explore rich physics in low-dimensional systems.

\end{abstract}

\maketitle

\renewcommand\linenumberfont{\normalfont\tiny}



\section{\label{sec:intro}Introduction}
Topological materials emerged as a fascinating class of materials in condensed matter physics, holding promise for applications such as spintronics and quantum computation.~\cite{he2019topological,he2022topological,vsmejkal2018topological} 
Recently, there has been a growing interest in the research of topological thermoelectrics, focusing on novel topological materials for thermoelectric energy conversion.~\cite{fu2020topological,xu2017topological}
For topological insulators, there are various properties desirable for thermoelectric applications: electronic energy bands with linear dispersions for high carrier mobility, low thermal conductivity due to heavy elements and lattice instability,~\cite{yue2020phonon} a small band gap due to strong spin-orbit coupling, and the topologically protected surface states that are stable against backscattering and defects.~\cite{fu2020topological,xu2017topological}
For topological semimetals, Skinner and Fu first theoretically proposed the non-saturating thermopower in a magnetic field.~\cite{skinner2018large}  
Experiments have confirmed that the Seebeck coefficient of topological semimetals can increase without saturation with magnetic field.~\cite{han2020quantized,zhang2020observation} 
It has also been suggested that, close to topological phase transitions of topological materials, the thermal conductivity is reduced owning to the Kohn-anomaly-induced phonon softening by first-principles simulations, Raman spectroscopy, X-ray scattering, and neutron scattering, opening up a possibility to further enhance the thermoelectric performance.~\cite{yue2019soft,yue2020phonon,nguyen2020topological}
The above properties of topological materials are promising features for improving the thermoelectric energy conversion efficiency.
So far, the electronic properties of topological insulators and semimetals have been extensively explored~\cite{yue2019soft,yue2020phonon}, but the thermal transport properties of topological materials are less investigated. 
In this work, we focus on lattice vibrations and thermal transport properties of \ce{BaMnSb2} and its potential thermoelectric applications.
The thermal transport properties of topological materials, have not been as thoroughly investigated as their electronic properties~\cite{yue2019soft,yue2020phonon}. 
By focusing on the lattice vibrations and thermal transport properties of \ce{BaMnSb2}, our study aims to deepen the understanding of heat conduction mechanisms in topological insulators and semimetals.

\ce{BaMnSb2} is an orthorhombic topological semimetal composed of alternating layers of Sb, Ba, and MnSb [see Fig.~\ref{fig:structure}(a) for structural illustration].
\ce{BaMnSb2} was previously assigned to a tetragonal structure with a Sb square net~\cite{cordier1977darstellung}. Based on the first-principles calculations of the Sb square net structure, Farhan \textit{et al.}~\cite{Farhan2014} predicted that the Dirac fermions exist. 
Liu \textit{et al.}~\cite{Liu2016} and Huang \textit{et al.}~\cite{Huang2017} reported the experimental discovery of \ce{BaMnSb2} hosting nearly massless fermions with high mobility and a nontrivial Berry phase.
However, recent studies identified a small orthorhombic distortion on the Sb layer, with space group $Imm2$ and $a = 4.4583$~\AA, $b =  4.5141$~\AA, $c=24.2161$~\AA.~\cite{Sakai2020} 
The distortion induces a zigzag chain-like structure on the Sb layer, opening a gap for the Dirac fermions and leading to spin-valley locking.~\cite{Sakai2020, Liu2021}
The signature of electronic band structure near the Fermi level is further confirmed by high-resolution angle-resolved photoemission spectroscopy (ARPES) measurements.~\cite{Rong2021}
In \ce{BaMnSb2}, the charge dynamics~\cite{Yoshizawa2022}, surface morphology~\cite{Zou2022}, pressure dependence~\cite{Yin2022}, and phonon helicity~\cite{Hu2021} have also been explored to reveal the rich physics of this material.
Recent experiments~\cite{Huang2020} have demonstrated that \ce{BaMnSb2} exhibits the low thermal conductivity and modest thermal power, making it a potential candidate for thermoelectric application. 
However, despite extensive experimental studies, a thorough computational investigation of lattice vibration of \ce{BaMnSb2} remain unexplored, and it is interesting to look into the thermal transport properties of this topological material because of its possible peculiar thermal transport behavior.

In this study, we aim to provide a state-of-the-art first-principles analysis of \ce{BaMnSb2} electronic structure, phononic structure, and thermoelectric transport properties, which are pivotal to understand and optimize the thermoelectric performance of this material.
To delve into the underlying physics of \ce{BaMnSb2} thermal transport properties, we employ density functional theory (DFT) to investigate the electronic and phononic structures, taking into account the on-site Coulomb repulsion for 3$d$ orbitals and spin-orbit coupling, which are important for this material. 
We reveal that the orthorhombic distortion of the Sb layer stabilizes the lattice of \ce{BaMnSb2} and opens a gap in the electronic bandstructure, providing another example of lattice instability near a topological phase transition~\cite{yue2020phonon}.  
Employing the anharmonic three-phonon scattering and semiclassical Boltzmann theory, we obtain the phonon and electron transport properties of \ce{BaMnSb2}, and explore the strategies to increase the thermoelectric figure of merit through alloying. 
The comprehensive study on the thermal transport properties of \ce{BaMnSb2} provides insights into the enhancement of thermal power, with potential applications in a variety of energy-efficient technologies.

This paper is organized as follows: Section \ref{sec:methods} outlines the computational methodology employed for our first-principles calculations, including DFT, phonon dispersion calculations, and electron transport properties. In Section \ref{sec:results}, we present and discuss our results, focusing on the electronic structure, ground states, phonon dispersion under various distortions, phonon and electron transport of \ce{BaMnSb2}. Section \ref{sec:conclusion} concludes this paper.

\section{\label{sec:methods}Methodology}
\subsection{Computational details}
First-principles calculations were performed in Vienna Ab initio Simulation Package (\code{VASP}) version 6.3.1~\cite{kresse1996vasp1, kresse1996vasp2} using projector augmented wave (PAW) potentials~\cite{blochl1994projector}. 
For Ba, the valence electrons consist of $5s^26p^66s^2$ orbitals; for Mn $3p^63d^54s^2$ orbitals; and for Sb $5s^25p^3$ orbitals.
A plane-wave basis set with a kinetic energy cutoff of 400 eV was utilized, and Gaussian smearing was applied with a broadening of 0.01 eV.  
For the unit cell, a Monkhorst-Pack~\cite{monkhorst1976special} 12$\times$12$\times$6 {\bf k}-point mesh is used for Brillouin zone integration.
Convergence criteria were set to $1\times10^{-8}$~eV for total energy and 0.001~eV/\AA~ for the Hellmann-Feynman force on each atom.
The Perdew–Burke–Ernzerhof (PBE) exchange–correlation functional was employed for calculations.~\cite{perdew1996generalized}
The on-site Coulomb interaction ($U$) on Mn 3$d$ orbitals as described by spherically averaged DFT+$U$ method~\cite{dudarev1998electron}, in which the Hamiltonian only depends on $U_\text{eff}$. $U_\text{eff}$  was set to 5 eV, which was determined using the linear response theory~\cite{cococcioni2005linear} and is in line with reported values in other studies~\cite{Farhan2014,mu2019influence}. 
The spin-orbit coupling (SOC) and PBE+$U$ are included for band structure and thermal calculations.~[see Fig.~S1(a) in Supplemental Material (SM) for the influence of SOC on the electronic band structure.\cite{SM}]
The PBE+$U$+SOC calculation employs a smaller \textbf{k} grid $\Gamma$-7$\times$7$\times$3 to compromise the computational expense.

\subsection{Phonon transport}\label{subsec:3rd}
The phonon properties of the material were calculated using the \code{Phonopy}~\cite{togo2015first} based on the Hellmann-Feynman forces~\cite{feynman1939forces}. 
The input interatomic force constants (IFCs) were obtained from \code{VASP} calculations. 
64-atom supercells with dimension of $2\times 2 \times 1$ are generated for finite displacement calculations, and the default atomic displacement amplitude (0.01~\AA) is applied. 
A Monkhorst-Park $4\times 4 \times 1$ \textbf{k}-mesh grid was used in the supercells to ensure the convergence of the phonon dispersion.

We calculated the lattice (phonon-contributed) thermal conductivity ${\kappa}_\text{ph}$ by iteratively solving the phonon Boltzmann transport equation using \code{ShengBTE} including the isotope-scattering effect.~\cite{li2014shengbte}
The second-order and third-order calculations with 5 neighbors are first performed with PBE+$U$ and PBE+$U$+SOC to show the independence of the SOC, as illustrated in Supplemental Material Fig.~S1\cite{SM}.
The 3rd order calculations employs 9 neighbors with 1756 finite-displacement jobs under PBE+$U$ functional. 
The convergence with respect to the number of neighbors is shown in Supplemental Material Fig.~S1(d)\cite{SM}. 
The supercell size and \textbf{k}-grid sampling of 3rd order calculations are consistent with the 2nd order calculations.
The broadening parameter in \code{ShengBTE} is chosen as 0.1 and \textbf{q}-mesh grid used is $10\times 10 \times 10$. 
The convergence test of the broadening parameter and \textbf{q}-mesh grid induces less than 0.05~$\mathrm{Wm^{-1}K^{-1}}$(3\% in in-plane direction and 6\% in z direction) difference in thermal conductivity.

\subsection{Electron transport}

We consider electron transport by linearized Boltzmann transport equation using \code{BoltzTraP}~\cite{Madsen2006,Madsen2018}, where the collision term is given by the constant relaxation time approximation.
The PBE+$U$+SOC electronic band structures obtained from the \code{VASP} calculations on a $\Gamma$-centered $62\times62\times10$ \textbf{k}-grid served as input for BoltzTraP. 
The Fermi surface is then Fourier interpolated~\cite{Madsen2006,Madsen2018} on a $\Gamma-201\times199\times38$ \textbf{k}-grid.
The output properties calculated using \code{BoltzTraP} in this work are thermopower and Hall carrier concentrations, which do not depend on the relaxation time within the constant relaxation time approximation.

\begin{figure*}[tbp]
\includegraphics[width=\textwidth]{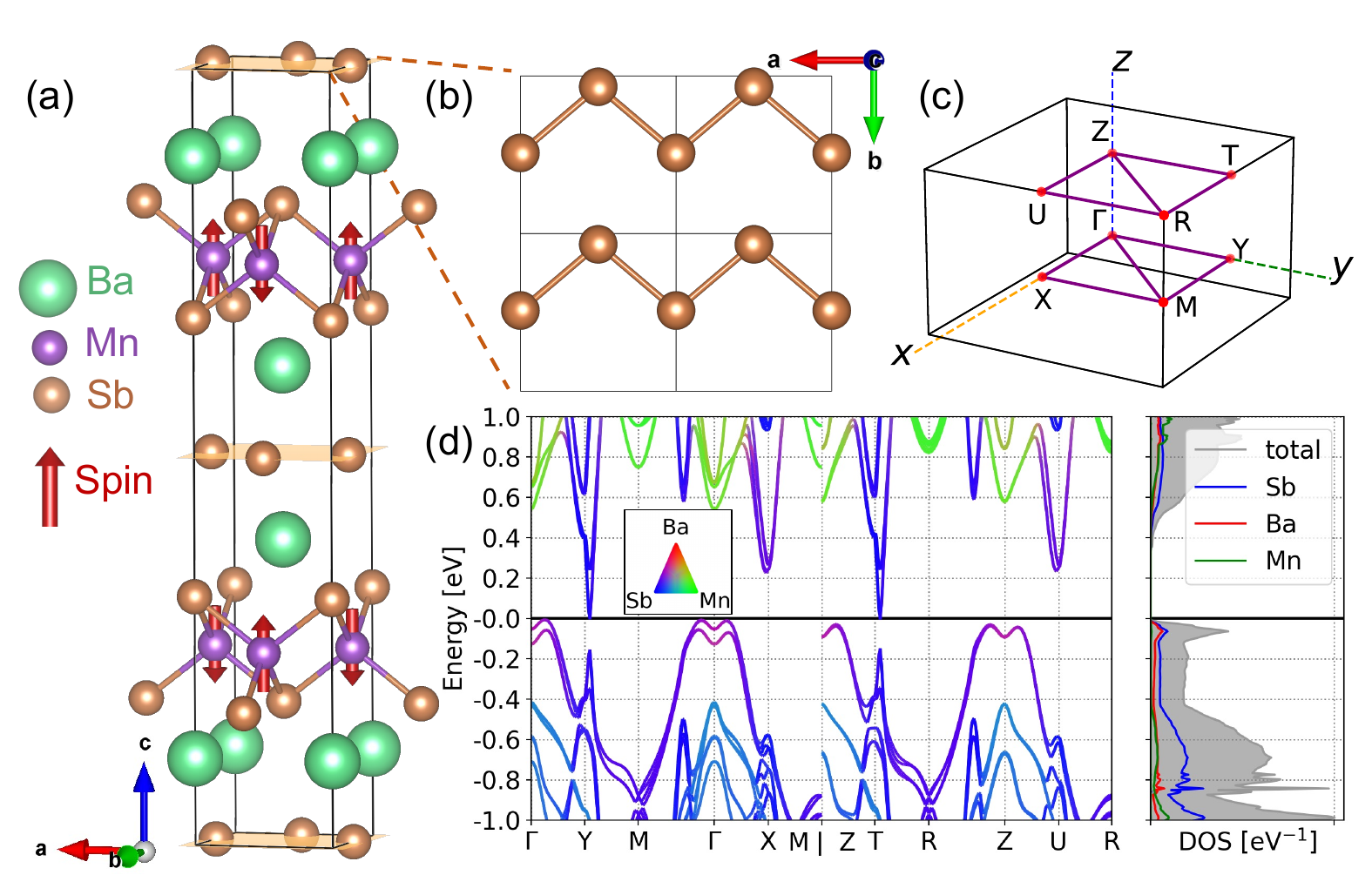}
\caption{\label{fig:structure} 
(a) The atomic structure of orthorhombic \ce{BaMnSb2}, comprised of alternative stacking of Ba, Sb, and MnSb layers.
The Mn atoms carry spin $S=5/2$ with a G-type antiferromagnetic spin structure. The red arrows denote spins. Green, purple and brown spheres denote Ba, Mn and Sb atoms, respectively.
(b) The zigzag structure of the Sb layer.
(c) The Brillouin zone of an orthorhombic structure with selected \textbf{k}-points.
The purple lines show two \textbf{k}-point paths  differing in $z$ direction: $\Gamma$-$Y$-$M$-$\Gamma$-$X$-$M$ and $Z$-$T$-$R$-$Z$-$U$-$R$.
(d) The band structure and element-projected density of states (DOS) of \ce{BaMnSb2} using PBE+$U$+SOC.
The color bar encodes atomic nature of the electron states.
}
\end{figure*}

\section{\label{sec:results}Results and Discussions}

\subsection{\label{subsec:ground}Ground state of \ce{BaMnSb2}}
In this section, we present the results of our first-principles calculations for the electronic ground state of \ce{BaMnSb2}.

The orthorhombic \ce{BaMnSb2} atomic structure is shown in Fig.~\ref{fig:structure}(a). 
\ce{BaMnSb2} is comprised of the alternative stacking of Ba, Sb, and MnSb layers.
The calculated lattice parameters are listed in Table~\ref{tab:spin}, which compare well with the experiment ones. Our DFT calculations using PBE slightly overestimate the lattice parameters by 2\%.
\ce{BaMnSb2} has a G-type antiferromagnetic ground state~\cite{Liu2016,Sakai2020} [see red arrows in Fig.~\ref{fig:structure} (a)], with S=$5/2$ on each Mn$^{2+}$ cation. Our calculated spin moment on Mn is 4.55 $\mu_B$, in a high spin state with a half-filled $d$ shell.
All spins are aligned along the easy-axis of $c$ axis.~\cite{Liu2016}  
To verify this, we calculated the magnetocrystalline anisotropy energy (MAE) by comparing the relativistic total energies (with SOC) for different spin quantization axis (see Tab.~\ref{tab:spin}). 
We do observe the lowest total energy with all spins aligned along $c$ axis. 
The calculated MAE is about 0.295~meV per formula unit (f.u.). 
This low MAE is consistent with the reduced spin orbit interaction for the case of half-filled $d$ shell (3$d^5$ for Mn cation) and the negligible orbital moment (0.01~$\mu_B$ from our calculation).

From the magnetic torque experiment, Huang \textit{et al.} declared a notable ferromagnetic spin canting in addition to the primary G-type antiferromagnetic order~\cite{Huang2017,Huang2020}, resulting in a secondary ferromagnetic order parameter. 
To explore the possible spin canting, we explicitly introduce various initial spin canting patterns and converge the charge density and spin density. 
The spins always relax back to the $z$ axis precisely, maintaining a G-type antiferromagnetic ordering. Our result does not support the existence of a weak ferromagnetism.

Half-filled $d$ shells are optimal for antiferromagnetic coupling. 
The measured N\'eel temperature of \ce{BaMnSb2} is 283-286 K~\cite{Liu2016,Sakai2020,Huang2017,Huang2020}. 
To estimate the magnetic transition temperature, we evaluated the exchange energy, which is defined as the energy separation between the magnetic ground state and the state in which the local moment on one Mn is flipped.  
In the Heisenberg model, the exchange energy E$_i$ on site $i$ is given by E$_i$ = 2 $\sum_j$ $J_{ij}$, where $J_{ij}$ is the Heisenberg exchange parameter. 
For the homogeneous system with equivalent magnetic sites, the mean-field T$_N$ for the classical Heisenberg model is equal to 1/6 of the exchange energy. 
Our calculated exchange energy for \ce{BaMnSb2} is 0.2028 eV, giving a classical mean field temperature of 392 K.  
The overestimation of T$_N$ is due to the neglect of spin correlation in a mean field model.  

Figure~\ref{fig:structure}(b) provides a top view of the Sb layer in Fig.~\ref{fig:structure}(a), revealing a zigzag chain-like structure in the Sb layer. The formation of zigzag distortion yields a 48.8~meV total energy reduction per formula unit [f.u.] compared to the high symmetry tetragonal structure. 
The zigzag distortion results in a shorter bond between the two Sb atoms, which are located at (0.5, 0.068, 0) and (0, 0.488, 0)  in Fig.~\ref{fig:structure}(b) [Note the coordinates are in the basis of lattice vectors of the unit cell].
The band structure of \ce{BaMnSb2} is depicted in Fig.~\ref{fig:structure}(d) and the high symmetry \textbf{k}-points are defined in Fig.~\ref{fig:structure}(c).
The two dimensional k-paths ($\Gamma$-$Y$-$M$-$\Gamma$-$X$-$M$ and $Z$-$T$-$R$-$Z$-$U$-$R$) have very similar band structures, which corresponds to the layered atomic structure in Fig.~\ref{fig:structure}(a).
It is worth noting that the conduction band minimum (CBM) near the Y point and the valence band maximum (VBM) near the $\Gamma$ point occur at the same energy. 
The CBM electron state is a topological state with its wavefunction originating from the Sb zigzag layers.
The CBM state is located at (0.049, 0.5, 0.5), while the VBM state is located at (0, 0.13, 0), both in the basis of reciprocal lattice vectors.
The VBM state is topologically trivial, which is much flatter than the steep topological CBM state.
The drastically different curvatures of CBM and VBM states lead to the distinct shape of the density of states (DOS) also shown in Fig.~\ref{fig:structure}(d).
The lowest CB state is steep and contributes to a tiny DOS, while the VBM DOS is much larger due to the less-dispersive band curvature, which yields a van Hove singularity.
This feature is important for \ce{BaMnSb2} as a potential candidate for thermoelectric applications, which will be discussed in Sec.~\ref{subsec:electron}.

\begin{table}[htb]
\caption{\label{tab:spin}%
The calculated lattice parameters of \ce{BaMnSb2} unit cell, compared to the experiment~\cite{Sakai2020}.
The total energies of the G-type antiferromagnetic states for different spin quantization axis using PBE+$U$+SOC. All energies are referenced to the case of spin quantization axis along $z$.
}
\begin{threeparttable}[b]
\begin{ruledtabular}
\begin{tabular}{cccccc}
Lattice parameter & $a$ & $b$ & $c$ 
\\
\colrule
Calc.~[\AA] & 4.5563 & 4.6120 & 24.8195
\\ \hline
Expt.~[\AA]~\cite{Sakai2020} & 4.4583 & 4.5141 & 24.2161
\\  
\hline \hline
Spin quantization axis & $x$-[100] & $y$-[010] & $z$-[001] 
\\
\colrule
Energy~[meV/f.u.] & 0.284 & 0.307 & 0.00
\end{tabular}
\end{ruledtabular}
\begin{tablenotes}
\item 
\end{tablenotes}
\end{threeparttable}
\end{table}


\subsection{\label{subsec:phonon}Phonon transport}
Now we move to the investigation of the phonon band structure of \ce{BaMnSb2}, which is pivotal to understand its lattice stability and thermal transport property. 
We first explore the importance of orthorhombic distortion in stabilizing the lattice and its impact on the electronic structure. 
We interpolate the structure between the tetragonal lattice (denoted as 0\% distortion) and the orthorhombic lattice (denoted as 100\% distortion) to acquire an intermediate configuration, which we refer to as 50\% distortion.
The phonon dispersion is calculated for the above three structures to explore the structural instability.

\begin{figure*}[!thp]
\includegraphics[width=\textwidth]{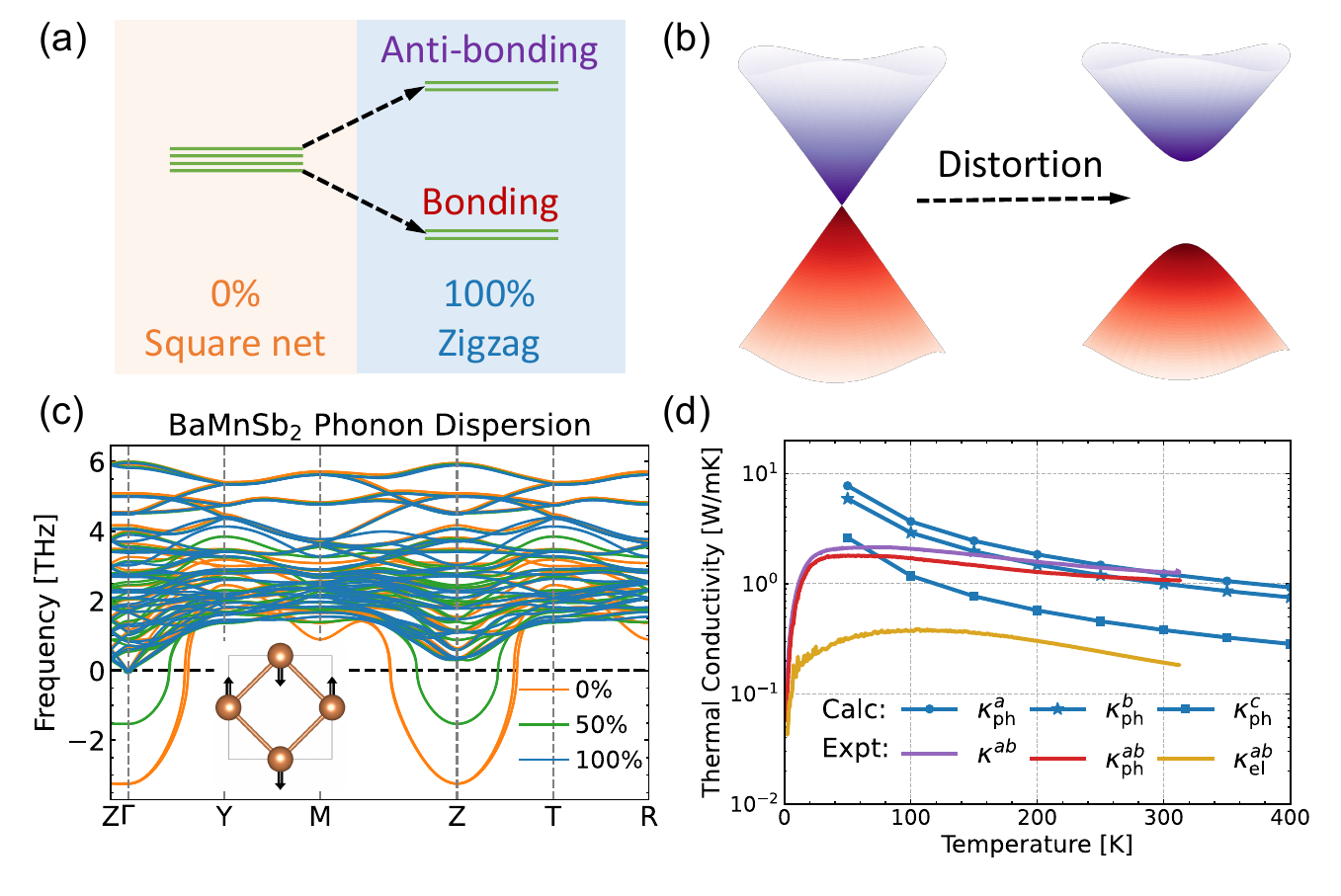}
\caption{\label{fig:phonon} 
(a) A sketch of band splitting into bonding-antibonding states, from the high-symmetry tetragonal Sb square net case to the low-symmetry orthorhombic Sb zigzag distortion case. 
(b) The band visualization of the avoid crossing for the bonding-antibonding splitting. 
(c) The phonon dispersion under 0\% , 50\% (intermediate), and, 100\% distortion.
0\% distortion exhibits imaginary phonon modes, and the magnitude of the imaginary modes gradually diminishes with larger distortion.
(d) The thermal conductivity $\kappa$ of \ce{BaMnSb2}, including electron-mediated $\kappa_\mathrm{el}$ and phonon-mediated $\kappa_\mathrm{ph}$ parts.
``Calc'' are calculated phonon thermal conductivity along $a,b,c$ directions.
``Expt'' are the experimental thermal conductivity in Ref.~\onlinecite{Huang2020}.
}
\end{figure*}

The phonon dispersions of 0\%, 50\%, and 100\% distortion structures are shown in Fig.~\ref{fig:phonon}(a).
The calculations are conducted using PBE+$U$ and we have verified that PBE+$U$+SOC yields the same phonon structure obtained from PBE+$U$ (see Supplemental Material Fig.~S1\cite{SM}).
From Fig.~\ref{fig:phonon}(a), the 0\% and 50\% distortion structures exhibit imaginary frequencies at both $\Gamma$ and $Z$ points. 
With a larger distortion, the magnitude of negative frequencies gradually diminishes.
The orthorhombic structure with 100\% distortion is stable without any imaginary-frequency phonon modes. 
By visualizing the eigenvector of the soft phonon mode for 0\% distortion at $\Gamma$ point (see inset of Fig.~\ref{fig:phonon}[c]), the vibrational movement is exclusively attributed to the two Sb atoms within the Sb layer. 
These two Sb atoms in Fig.~\ref{fig:structure}(b) vibrate against each other along the $b$-direction, thereby forming the zigzag distortion in the Sb layer.

To elucidate the formation mechanism of the zigzag structure, Tremel and Hoffman~\cite{tremel1987square} conducted an in-depth investigation into layered stacking structures with square atomic nets to discern the pivotal element triggering the distortion.
Their argument lies in whether the electron filling is close to the band crossing point in the high-symmetry square atomic net structure, as shown in Fig.~\ref{fig:phonon}(b).
Although such a distortion from a high-symmetry configuration induces an increase in the elastic energy, it also creates a bonding/anti-bonding splitting from the degenerate orbitals in Fig.~\ref{fig:phonon}(a). 
With the correct electron filling - that is, the Fermi level lying close to the Dirac crossing - only the bonding states become occupied, leading to a reduction in the electron energy. 
This thereby gives rise to the preference for a symmetry-lowering distortion. This argument is in the same spirit as the Peierls transition.\cite{peierls1955quantum}
The recent discovery of the distorted \ce{BaMnSb2} structure revealed its proximity to the non-distorted structure, which is unstable.
Hence, as displayed in Fig.~\ref{fig:phonon}(c), the phonon dispersion for unstable structures exhibits optical phonon modes with imaginary frequencies, and only for the fully distorted orthorhombic structure, the frequencies of those optical modes become positive.
This presents an additional perspective on the lattice instability observed near the topological phase transition, as discussed in a few previous studies.~\cite{nguyen2020topological,yue2019soft,yue2020phonon}

We further calculate the thermal conductivity of \ce{BaMnSb2} using \code{ShengBTE}. The calculated temperature-dependent thermal conductivity of the lattice along different crystallographic axis ($\kappa^a_\text{ph}$, $\kappa^b_\text{ph}$, $\kappa^c_\text{ph}$) are illustrated in Fig.~\ref{fig:phonon}(d). 
The experimentally measured in-plane thermal conductivity~\cite{Huang2020} is also shown for comparison. Note that the experimental phonon thermal conductivity is derived by subtracting the electron contribution $\kappa_\mathrm{e}^{ab}$ from the total thermal conductivity $\kappa^{ab}$. 
At low temperatures, the discrepancy in lattice thermal conductivity between the calculation and the experiment is due to the omission of the phonon-boundary scattering (the dominant scattering mechanism at low T) in the calculations.~\cite{chen2005nanoscale}
At elevated temperatures, the phonon thermal conductivity is mainly limited by phonon-phonon scattering. 
We observe a good agreement with experimental lattice thermal conductivity near the room temperature.
At $300$~K, the calculated phonon thermal conductivities are $\kappa_\mathrm{ph}^a$=1.23, $\kappa_\mathrm{ph}^b$=1.00, and $\kappa_\mathrm{ph}^c$=0.38~$\mathrm{Wm^{-1}K^{-1}}$. 
At temperatures above 300K, the calculated thermal conductivity trend falls below the experimental one, and is probably due to the wave-like phonon-tunneling thermal conductivity.~\cite{di2023crossover}
The higher in-plane thermal conductivity than the out-of-plane thermal conductivity is consistent with the structural feature of these layered materials, in which the in-plane bonding is stronger than the out-of-plane bonding. 
There is a small in-plane anisotropy between $\kappa_\mathrm{ph}^a$ and $\kappa_\mathrm{ph}^b$ due to the orthorhombic distortion. The calculated $\kappa_\mathrm{ph}^a$ and $\kappa_\mathrm{ph}^b$ are fairly close to the experimental in-plane lattice thermal conductivity $\kappa_\mathrm{ph}^{ab}$=1.08~$\mathrm{Wm^{-1}K^{-1}}$.

Overall, the lattice thermal conductivities are rather low. This is due to the scattering of acoustic phonons with 
the low-frequency, or ``soft'', optical modes.
Further details regarding three-phonon scattering rates and cumulative thermal conductivity can be found in Fig.~S2 of the Supplemental Material~\cite{SM}.
While our investigation of the phonon transport properties affirms the potential of \ce{BaMnSb2} as a promising thermoelectric material due to its low lattice thermal conductivity, 
a complete understanding of its thermoelectric performance requires a simultaneous examination of the electron transport properties, which will be the focus of the subsequent section.


\begin{figure*}[htbp]
\includegraphics[width=\textwidth]{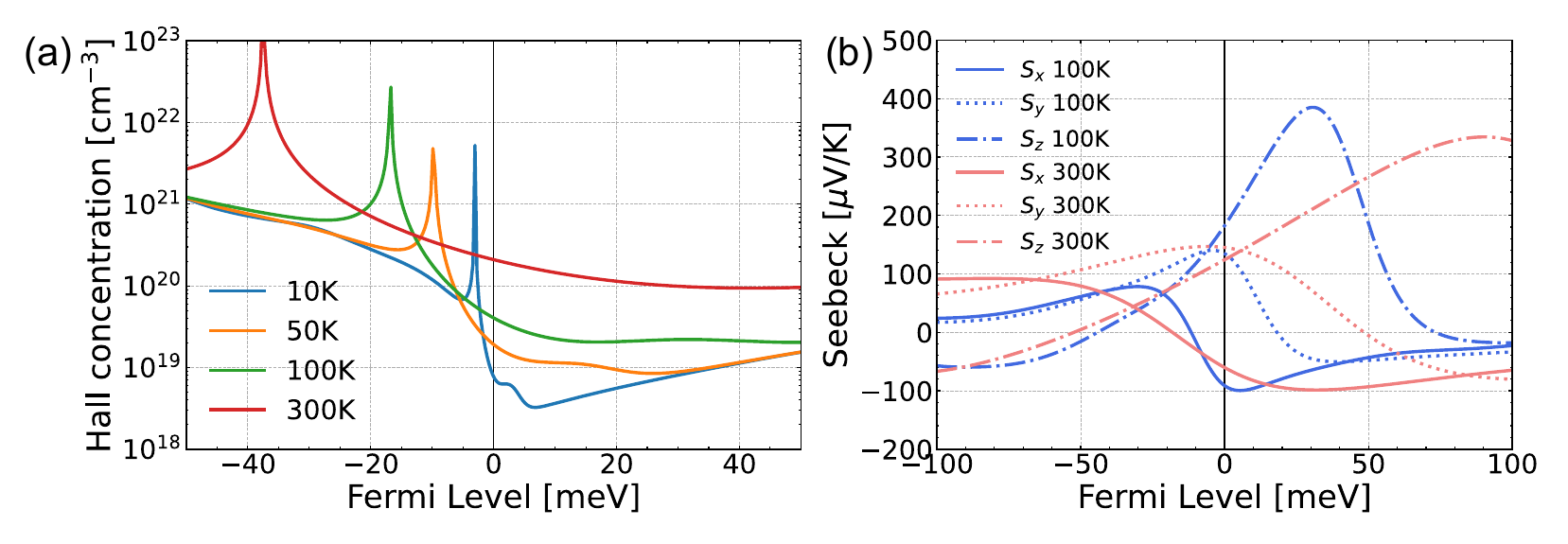}
\caption{\label{fig:electron} 
Electron transport properties of \ce{BaMnSb2} with constant relaxation time approximation. 
(a) Hall carrier concentration as a function of Fermi level.
(b) The Seebeck coefficient $S_x, S_y, S_z$ as a function of Fermi level
at temperatures of 100K and 300K.
}
\end{figure*}

\subsection{\label{subsec:electron}Electron transport}

We now focus on another critical facet of its thermoelectric performance --- electron transport.
To gain insight into electron transport, we first calculate an accurate electronic structure incorporating SOC (see Fig.~\ref{fig:structure}(d)) and employ the linearized Boltzmann transport equation to assess the transport property. The constant relaxation time approximation (RTA) is adopted in the process.

Figure~\ref{fig:electron}(a) shows the Hall carrier concentration as a function of the Fermi level ($E_F$) at different temperatures. $E_F$ $= 0$ eV corresponds to the charge neutrality condition for the pristine structure.
For positive $E_F$, the charge carriers are electrons, while for negative $E_F$, the charge carriers are holes. 
The singularity in the Hall concentration in the negative $E_F$ region is due to the dominant DOS of the less-dispersive top valence bands
The position of $E_F$ is sample-dependent and is related to the carrier concentration. 
In order to compare our calculation to experimental measurements~\cite{Liu2016,Liu2021,Sakai2020,Huang2017,Huang2020,Yin2022}, we estimate the actual position of $E_F$ by calculating the angle-dependent oscillation frequency and comparing it with the experimental Shubnikov-de Haas (SdH) measurement. 
The measured oscillation frequency~\cite{Huang2017,Liu2016} at each angle was reproduced from our DFT calculations, only when $E_F$ is set to 35 $~meV$. (see a detailed analysis in Supplemental Material Fig.~S4\cite{SM})
At this determined $E_F$, the Hall carrier concentration is about $1.0\times 10^{19}~\mathrm{cm}^{-3}$ at 10~K in Fig.~\ref{fig:electron}(a). 
It is close to the previously reported experimental values at 10~K: $1.16\times 10^{19} ~\mathrm{cm}^{-3}$ in Ref.~\onlinecite{Huang2017} 
and $1.4\times 10^{19} ~\mathrm{cm}^{-3}$ in Ref.~\onlinecite{Liu2016}, respectively.

Figure~\ref{fig:electron}(b) illustrates the Seebeck coefficient as a function of the Fermi level at 100 K and 300 K. The Seebeck coefficients along three crystallographic axes are calculated separately. 
In general, we anticipate a positive Seebeck coefficient ($S>0$) for hole carriers ($E_F<0$) and a negative $S$ for electron carriers ($E_F >0$). 
At both temperatures ($T=100$~K and $T=300$~K) and for a positive $E_F$, \ce{BaMnSb2} exhibits an intriguing goniopolar behavior\cite{uchida2022thermoelectrics,wang2020anisotropic,radha2020topological}, namely, the material simultaneously possesses $p$-type conduction along $z$ direction ($S_z>0$) and $n$-type conduction along specific in-plane directions ($S_x<0, S_y<0$).
This unconventional behavior can be explained by the drastically different curvatures of the electronic bands near CBM and VBM states. 
As described in Sec.~\ref{subsec:ground}, the top of the valence band is very flat and almost dispersion-less while the bottom of the CB is much sharper, leading to a large contrast in the electron and hole DOS near $E_F$. 
At higher temperatures, the Fermi-Dirac distribution broadens near the Fermi level. 
Given that the hole carriers have a substantially larger DOS compared to that of the electrons~(see Fig.~\ref{fig:structure}d), the holes dominate the contribution to the Seebeck coefficient, rendering a positive Seebeck even for a Fermi level deep into the conduction band.
More detailed analysis on the underlying mechanism for this distinctive behavior is elaborated in Fig.~S3 of the Supplemental Material~\cite{SM}. 
This intriguing goniopolar behavior opens avenues for innovative electronic devices that leverage direction-dependent carrier polarity, such as thermoelectric converters, photovoltaic cells, transistors, and sensors.\cite{wu2022thermoelectric,wu2021thermoelectric,chenni2007detailed,jordehi2016parameter,liu2021promises,vetelino2017introduction} 
Beyond this, goniopolar materials can showcase complex transport phenomena unattainable in conventional semiconductors, including anisotropic magnetoresistance, the Nernst effect, and the Hall effect. 
With such unique features, goniopolar materials carry immense potential to transform sectors like electronics, energy conversion, and information technology.

Bringing together the unique electron transport behavior and the low lattice thermal conductivity, we can estimate the thermoelectric figure of merit and evaluate the potential of \ce{BaMnSb2} as a thermoelectric material.
To avoid the complexity of subjectively choosing an electron relaxation time, we estimate it from experimental values.
The thermoelectric figure of merit is defined as $ZT=\frac{\sigma S^2 T}{\kappa}$, where $S$ is the Seebeck coefficient, $T$ the temperature, $\sigma$ the electrical conductivity, and $\kappa$ the thermal conductivity.~\cite{chen2005nanoscale}
$\sigma$ and $\kappa$ are from experimental measurements and $S$ is based on our DFT calculations. 
At room temperature (300~K), the variables are reported as: 
The in-plane thermal conductivity $\kappa=1.3$~W/mK~\cite{Huang2017}; 
The in-plane Seebeck coefficient $72~\mu \mathrm{V/K}$~\cite{Huang2017} (within the range of Fig.~\ref{fig:electron}(b)); 
The in-plane electrical conductivity $\sigma=3.7\times 10^4 ~\mathrm{S/m}$~\cite{Huang2017} 
or $2\times 10^5 ~\mathrm{S/m}$~\cite{Liu2021}.
Accordingly, the thermoelectric figure of merit $ZT$ of the \ce{BaMnSb2} material is either 0.03~\cite{Huang2017} or 0.24~\cite{Liu2021} at 300~K.

The thermoelectric performance can potentially be further enhanced by adjusting the temperature and doping levels to optimize the figure of merit.
Additionally, other standard techniques, such as grain refinement and modulation doping may also improve the figure of merit.~\cite{zhou2018routes,shi2020advanced,zebarjadi2011power}
In light of these insights into the unique phonon and electron transport properties of \ce{BaMnSb2}, we foresee its potential to impact the design and experimental pursuits in thermoelectric materials. 
Our findings will be informative in advancing the efficient and sustainable energy solutions across diverse scientific disciplines.

\section{\label{sec:conclusion}Conclusions}
In summary, we present an state-of-the-art first-principles study of thermal transport in the Weyl semimetal \ce{BaMnSb2}. 
We use density functional theory to comprehensively investigate the electronic structure, phononic structure, and electron transport of \ce{BaMnSb2}, taking into account the effects of Coulomb repulsion and spin-orbit coupling. 
Our calculations reveal a low lattice-mediated thermal conductivity of \ce{BaMnSb2} due to the phonon scattering by low-frequency optical phonon modes when assessing three-phonon scattering. 
The electron transport exhibits an unconventional goniopolar behavior of the Seebeck coefficient as a function of the Fermi level.
These findings, along with the estimated figure of merit, render \ce{BaMnSb2} as a potential thermoelectric material for an efficient and sustainable energy solution. 
We anticipate that our insights will be informative for the design and experimental realizations of thermoelectric materials towards improved performance.


\begin{acknowledgments}
The work was supported by the program of UCSB Quantum Foundry, which is supported by the National Science Foundation under Grant DMR-1906325. Sai Mu would like to acknowledge the startup fund from the University of South Carolina and an ASPIRE grant from
the VPR’s office at the University of South Carolina.
The computing resources were provided by National Energy Research Scientific Computing Center (NERSC), a U.S. Department of Energy Office of Science User Facility located at Lawrence Berkeley National Laboratory, operated under Contract No. DE-AC02-05CH11231. 
\end{acknowledgments}



%

\end{document}



\title{Supplementary Information }

\author{Yubi Chen}
\affiliation{%
 Department of Physics, University of California, Santa Barbara, California 93106-9530, USA}%
 \affiliation{%
Department of Mechanical Engineering, University of California, Santa Barbara, CA 93106-5070, USA}%
\author{Rongying Jin}
\affiliation{%
 SmartState Center for Experimental Nanoscale Physics, Department of Physics and Astronomy, University of South Carolina, SC 29208, USA}%
\author{Bolin Liao}
\email{bliao@ucsb.edu}
\affiliation{%
Department of Mechanical Engineering, University of California, Santa Barbara, CA 93106-5070, USA}%
\author{Sai Mu}
\email{mus@mailbox.sc.edu}
\affiliation{%
 Center for Experimental Nanoscale Physics, Department of Physics and Astronomy, University of South Carolina, SC 29208, USA}%


\maketitle






\section{Computational setup for electron and thermal properties}

\begin{figure*}[htbp]
\includegraphics[width=\textwidth]{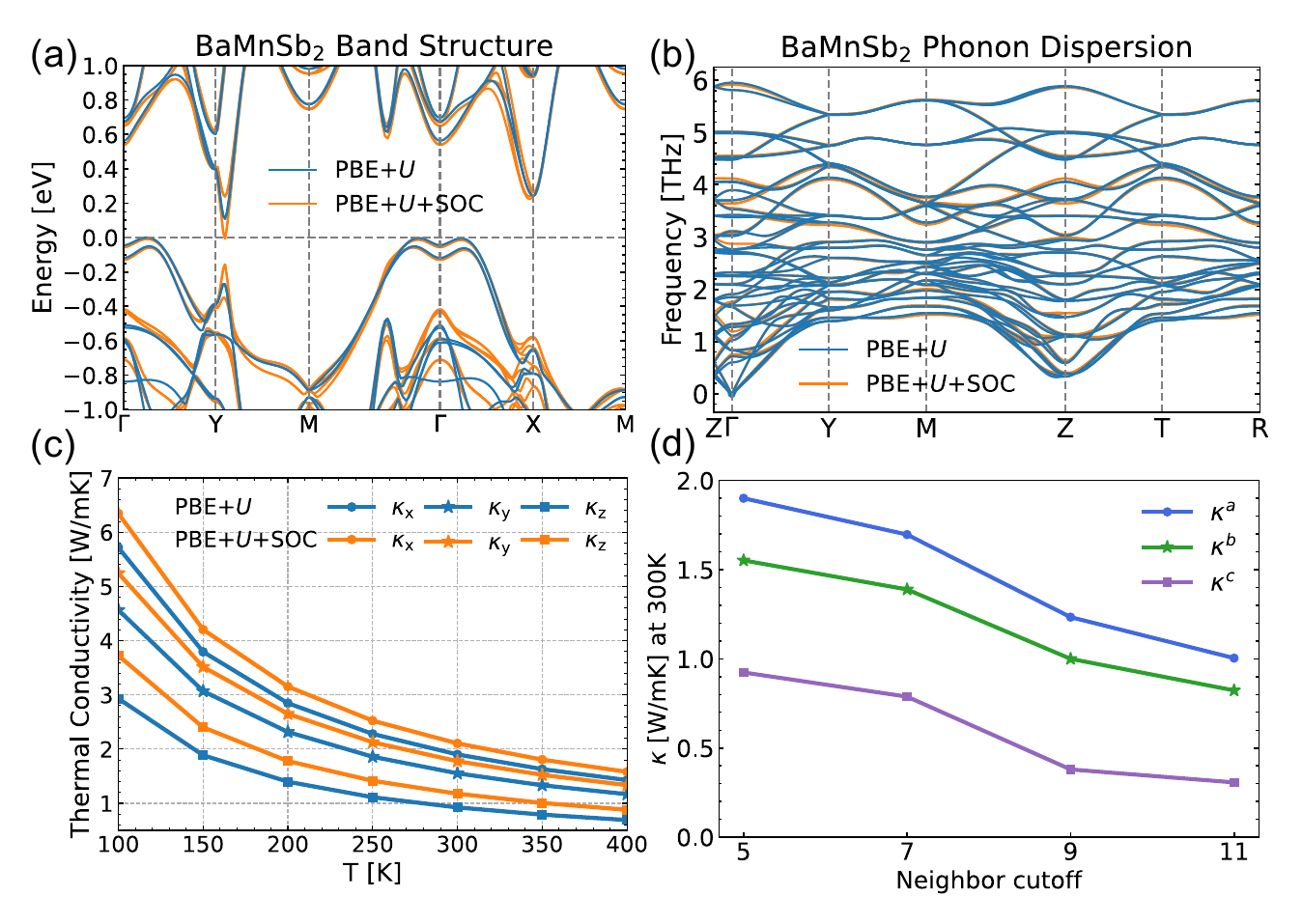}
\caption{\label{figs:pbeU} 
(a) Electron band structure of \ce{BaMnSb2} for PBE+$U$ and PBE+$U$+SOC.
(b) The phonon dispersion with and without SOC calculated by \code{Phonopy}.
(c) The lattice thermal conductivity with and without SOC computed by \code{ShengBTE}.
(d) The thermal conductivity convergence test with respect to the neighbor cutoff in \code{thirdorder} expansion.
}
\end{figure*}


The three computational setups (PBE+$U$, and PBE+$U$+SOC) are described by in the Methodology section of main text.
Figure~\ref{figs:pbeU}(a) reveals the electron band structure comparison for PBE+$U$ and PBE+$U$+SOC. 
The effects of SOC is mainly on the topological states on the Sb layer.
In general, the band structures resemble each other, so we expect small difference in thermal property calculations.
Figure~\ref{figs:pbeU}(b) and (c) shows the phonon dispersion and lattice thermal conductivity with and without SOC.
The lattice thermal conductivity has 724 finite-displacement jobs under five-neighbor cutoff.
The negligible effect of SOC on phonon dispersion and thermal conductivity shows that PBE+$U$ is good enough to capture the thermal properties of \ce{BaMnSb2}.
Figure~\ref{figs:pbeU}(d) shows the convergence of thermal conductivity in different directions ($\kappa^a, \kappa^b$ and $\kappa^c$) with respect to the number of neighbors in the third order cutoff.
With nine neighbors, the cutoff radius is 4.6~\AA, and further increasing neighbor cutoff has less than 0.25~$\mathrm{Wm^{-1}K^{-1}}$ difference in the thermal conductivity.

\begin{figure*}[htbp]
\includegraphics[width=\textwidth]{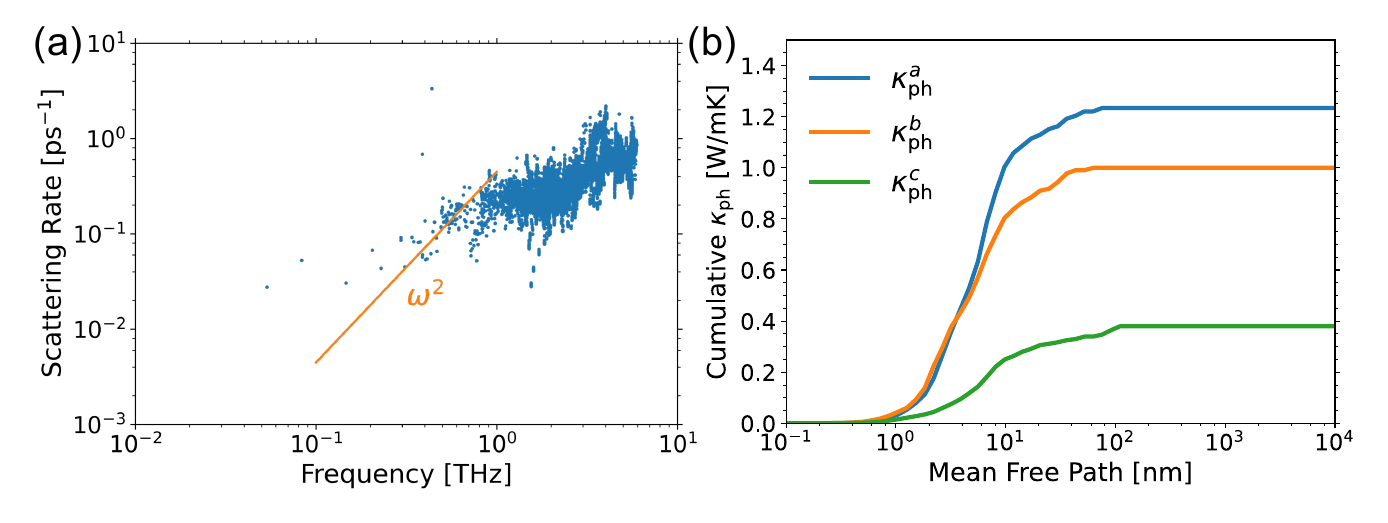}
\caption{\label{figs:thermal} 
(a) Three-phonon scattering rate of \ce{BaMnSb2} at 300~K with the frequency-squared trend indicated by the orange $\omega^2$ line.
(b) The cumulative phonon thermal conductivity $\kappa_\mathrm{ph}$ as a function of mean free path at 300~K.
}
\end{figure*}

Figure~\ref{figs:thermal}(a) shows the scattering rate at 300~K obtained from \code{ShengBTE}. 
It is generally expected that, for low-frequency acoustic phonons, the scattering rate exhibits the frequency-squared trend, as delineated by the orange $\omega^2$ line.~\cite{ziman2001electrons}
The low-frequency scattering rates are higher than this trend, meaning the optical phonons strongly scatter the acoustic phonons, greatly reducing the lattice thermal conductivity.
As a result, \ce{BaMnSb2} has a remarkably low lattice thermal conductivity, given a simple crystalline structure.

Figure~\ref{figs:thermal}(b) displays the size dependence of phonon thermal conductivity using \code{ShengBTE} by plotting the cumulative phonon thermal conductivity as a function of the mean free path at 300~K.
The accumulation function of $\kappa_\mathrm{ph}$ implies how phonons with different mean-free-paths contribute to the overall phonon thermal conductivity.
For \ce{BaMnSb2}, its thermal conductivity is predominantly contributed by phonons with the mean free path around 10~nm, and reaches saturation when the sample size is roughly 100~nm.
Using the phonon mean free path, we can identify which phonon-scattering mechanism, like phonon-phonon or phonon-boundary, becomes dominant for heat conduction at different size scale.

\begin{figure*}[htbp]
\includegraphics[width=\textwidth]{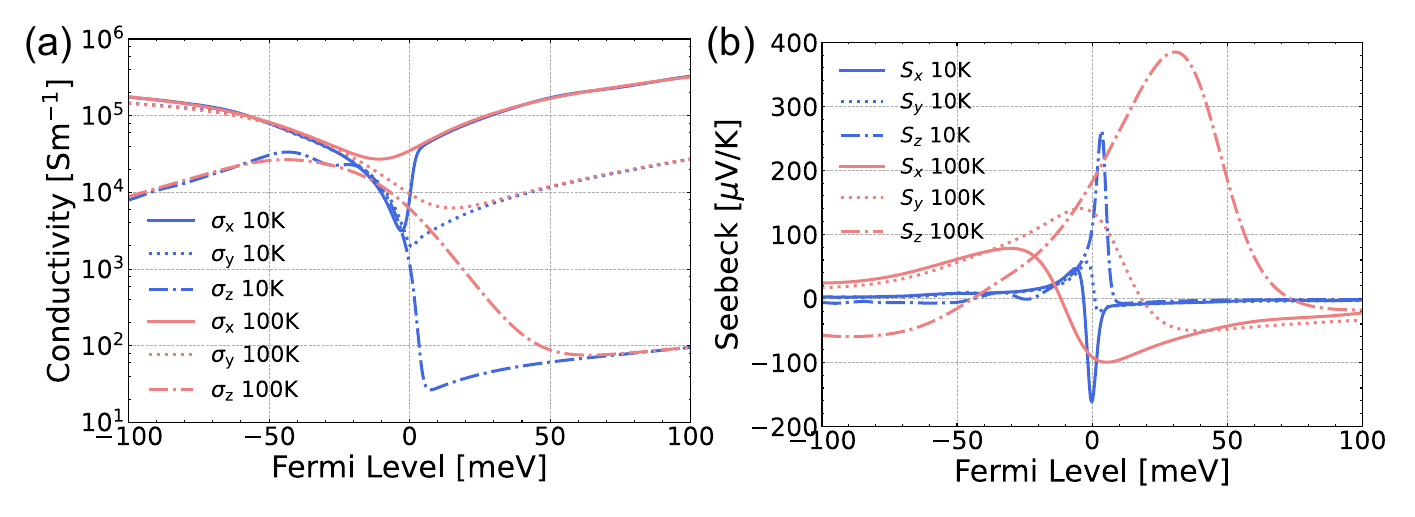}
\caption{\label{figs:btp2} 
(a) Electrical conductivity calculated by \code{BoltzTraP} with constant relaxation time $\tau=1\times 10^{14}$~s.
(b) Seebeck coefficient that is in the Fig.~3 of the main text.
The negative slope of electrical conductivity corresponds to the sign of the Seebeck coefficient.
}
\end{figure*}

The Seebeck coefficient is calculated by the Mott formula, in which $S(E_F)\sim -\frac{\sigma^\prime(E_F)}{\sigma(E_F)} $.~\cite{ashcroft2022solid}
$\sigma(E_F)$ is the electrical conductivity at Fermi level $E_F$.
Figure \ref{figs:btp2}(a) shows the numerically calculated $\sigma(E_F)$ with the constant relaxation time $\tau=1\times 10^{14}$~s.
In the conduction band $E_F>0$, we have $\sigma_x > \sigma_y > \sigma_z$ because the electron group velocity $v=\frac{\partial \varepsilon}{\partial k}$ evaluated in the electron band structure is the largest in $x$ direction $(v_x > v_y > v_z)$.
Even though, the density of states for conduction band is super low, the large group velocity $v_x$ brings in a steep increase, inducing a comparable $\sigma_x$ for conduction band and valence band with constant $\tau$.
For the $z$ direction, $\sigma_z$ is drastically different for $E_F<0$ and $E_F>0$ due to the tiny $v_z$.

In Fig.~\ref{figs:btp2}(b), the sign of the Seebeck coefficient corresponds to the negative slope of $\sigma(E_F)$.
At low temperature (10~K), the change of slope happens near the neutral point $E_F=0$. 
At high temperature (100~K), $\sigma(E_F)$ is smeared, and the slope change depends on the relative difference of $\sigma(E_F<0)$ and $\sigma(E_F>0)$.
For $\sigma_z$, the large slope gives a large positive $S_z$, which is hole type conduction. 
For in-plane $\sigma_x$ and $\sigma_y$, the higher electron mobility results in smaller $\sigma$ difference, and the Seebeck positive-to-negative transition is still close to the neutral point $E_F=0$.
Therefore, the in-plane conduction is electron type, and \ce{BaMnSb2} has applications as a goniopolar material.

\section{Quantum oscillation for Fermi level estimation}

\begin{figure*}[htbp]
\includegraphics[width=\textwidth]{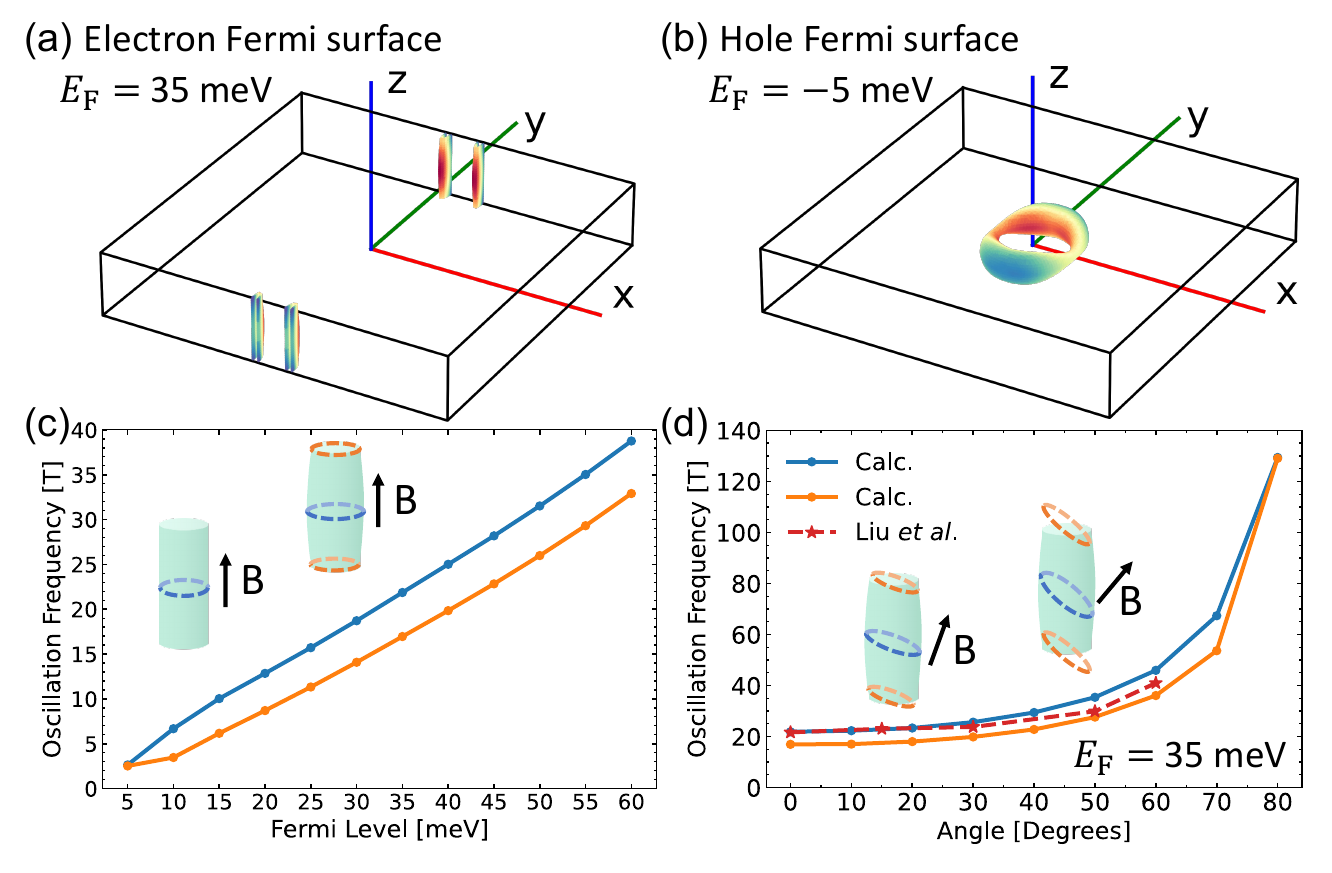}
\caption{\label{figs:oscillation} 
(a) Electron Fermi surface at Fermi level 35~meV above the neutral point.
(b) Hole Fermi surface at Fermi level $-$5~meV below the neutral point.
(c) Quantum oscillation frequency at a function of Fermi level. The small positive Fermi level has a cylinder shape, while larger Fermi level has a barrel shape with a bulge at the center.
The barrel-shape Feri surface has two extremal orbitals perpendicular to magnetic field direction $B$, giving rise to the larger blue and smaller orange areas or frequencies.
(d) The comparison of computed and experimental quantum oscillation frequencies.
The computational data varies with magnetic field rotating from $z$ to $x$ direction.
The two barrel shapes describe how the extremal orbitals evolve with different magnetic field direction.
The experimental data is from the Shubnikov-de Haas (SdH) oscillations of Ref.~\onlinecite{Liu2016}.
}
\end{figure*}

Quantum oscillation measurements are useful tools to determine the Fermi surfaces of the material.~\cite{ashcroft2022solid}
For metallic systems, applying a strong magnetic field (B) will lead to the Landau level quantization of electron orbitals, and thus the carrier density of states oscillates with the B field, and oscillations manifest in many experimental observations such as the Shubnikov-de Haas (SdH) oscillations in Ref.~\onlinecite{Liu2016}.
The oscillation frequency $F$ is closely related to the shape of Fermi surface by the Onsager relation $F=\frac{\hbar}{2\pi e} A$. 
$F$ is proportional to extremal areas $A$ of the Fermi surface perpendicular to the applied magnetic field, where $\hbar$ is the reduced Planck's constant and $e$ is the electron charge.

We can evaluate the oscillation frequency by a fortran-based package \code{SKEAF}~\cite{Rourke2012}. 
With a particular shape of Fermi surface like Fig.~\ref{figs:oscillation}(a) and (b), slices of Fermi surface perpendicular to the B field are first generated, and the areas of the Fermi surface slices are computed to find the extremal values.
Figure~\ref{figs:oscillation}(c) shows the calculated oscillation frequencies as a function of Fermi level.
The inside cylinder and barrel are schematic illustrations of the Fermi surface with smaller and larger Fermi levels. 
With magnetic field in $z$ direction, the extremal orbitals are the blue and orange dashed circles.
The blue (orange) electron orbital has a larger (smaller) area, and thus corresponds to a larger (smaller) oscillation frequency in the blue (orange) line.
Figure~\ref{figs:oscillation}(d) plots the oscillation frequency with rotating B field direction from $z$ to $x$.
With a larger tilting angle, the extremal orbitals have larger areas. 
We have found the oscillation frequency near Fermi level 35~meV agree reasonably well with the experimental data\cite{Liu2016}.
Hence, the Fermi level position is estimated to be 35~meV.

%